\documentstyle[12pt]{article}
\def\prd#1{{\em Phys.~Rev.}~{\bf D#1},\ }

\def\plett#1{{\em Phys.~Lett.}~{\bf #1B},\ }
\def\np#1{{\em Nucl.~Phys.}~{\bf B#1},\ }
\def\deg{\ifmmode{^{\circ}}\else ${^{\circ}}$\fi}
\def\pri{^{\,\prime}}

\newcommand{\gsim}{\,\raisebox{-0.13cm}{$\stackrel{\textstyle
        >}{\textstyle\sim}$}\,}
\newcommand{\lsim}{\,\raisebox{-0.13cm}{$\stackrel{\textstyle
        <}{\textstyle\sim}$}\,}
\newcommand{\bi}{\begin{itemize}}
\newcommand{\ei}{\end{itemize}}
\def\ed{\end{document}}
\def\be{\begin{equation}}
\def\ee{\end{equation}}
\def\vev#1{\left<{#1}\right>}
\def\lam{\ifmmode{\Lambda}\else $\Lambda$\fi}

\def\mp{m_{Pl}}

\def\quarter{\frac{1}{4}}
\def\thalf{{\textstyle\frac{1}{2}}}
\def\tquarter{{\textstyle\frac{1}{4}}}

\def\nl{N_{\lambda}}
\def\tev{\ \mbox{TeV}}
\def\pri{^{\, \prime}}

\def\ms{m_{SUSY}}
\def\msig{m_{\sigma}}
\def\msigt{\msig^2}
\def\lsig{\lambda_{\sigma}}
\def\lg{\lambda_G}
\def\msigp{m_{\sigma\pri}}

\def\fmunui{F_{\mu\nu}^i}
\def\ffmunui{F^{\mu\nu}_i}
\def\sighg{\sigma\mbox{--}G}
\def\mg{m_G}
\def\mgt{m_G^2}

\def\sigg{\sigma G}
\def\hedd#1{\noindent{\large\bf #1}\medskip}
\def\vzero{V^{(0)}}
\def\eb{\end{thebibliography}}
\def\nl{\newline}
\begin{document}
\begin{titlepage}
\begin{flushright}
{\sl NUB-3083/93-Th}\\
{\sl January 1994}\\
hepph/9402300
\end{flushright}
\vskip 0.5in
\begin{center}
{\Large\bf The Mass of the Dilaton }\\[.5in]
{Haim Goldberg}\\[.1in]
{\sl Department of Physics}\\
{\sl Northeastern University}\\
{\sl Boston, MA 02115}
\end{center}
\vskip 0.4in
\begin{abstract} It is shown that, in a
theory where the dilaton is coupled to a
Yang-Mills gauge field which enters a confining phase at scale \lam,\ the
dilaton may grow a mass $m_{dilaton}\sim \Lambda^2/\mp\sim
(\ms^{\;2}\mp)^{1/3}\sim 10^8\ \mbox{GeV}. $ This allows ample time for decay
before the electroweak era if
$\ms\simeq 1\ \mbox{TeV},$
and circumvents cosmological problems normally associated with its
existence.
\end{abstract}
\end{titlepage}
\setcounter{page}{2}
\hedd{Introduction}

The dilaton is a spin-0 field which is associated with the graviton and with
an axion as a massless excitation in string theory \cite{witten}. In the field
theoretic limit it persists as a massless spin-0 field in the no-scale
supergravity theory.  A truly massles dilaton is of course incompatible with
the success of Newtonian gravity, since it couples as a full strength
Brans-Dicke field. However, even if a dynamical mechanism were to be
found to stabilize the dilaton potential \cite{dilpot}, perhaps generating a
dilaton mass of the order of the the SUSY-breaking
$\ms\sim 1\tev,$ severe cosmological problems would still persist. Most
notably, the late decay time would cause a substantial reheating well after the
nucleosynthesis era \cite{dilatondecay}, vitiating a major success of the
standard cosmological model.
All of this could conceivably be modified if the dilaton were coupled to a
field whose interactions are strong: assuming that the dilaton masslessness is
not protected by a strict gauge principle, then mass growth could occur as a
result of mixing with composite spin-0 objects in the strongly-interacting
sector. This is the mechanism I will consider in this work.\bigskip

\hedd{Effective Potential}

I begin by presenting the minimal components of the
model necessary for  this work. I assume that the dilaton field $S$ is
universally coupled to all (hidden and unhidden) gauge fields
$\fmunui,$  giving a
contribution to the Lagrangian
\be
{\cal L}_{SF}= S\sum_i\fmunui\ffmunui\ \ .
\label{eq:sf}
\ee
The dilaton field is assumed to develop a vev $\vev{S}=1/g^2,$ leaving an
interaction with the remaining (quantum) field
\be
{\cal L}_{\sigma F}=\frac{1}{\mp}\sigma\sum_i\fmunui\ffmunui\ \ ,
\label{eq:sigf}
\ee
where $\sigma=S-\vev{S},$ and the gauge fields in Eq.~(\ref{eq:sigf}) are now
normalized with canonical kinetic energy. I now assume that at some scale
\lam\ the gauge coupling becomes strong. What I have mostly in mind, of course,
is the hidden-gaugino condensate mechanism for generating
SUSY-breaking \cite{nillesetal}. In such models, one commonly obtains

\be
m_{gravitino}\approx\ms \approx \frac{\lam^3}{\mp^2}\ \ ,
\label{eq:msig}
\ee
so that $\lam\sim 10^{14}\mbox{GeV}.$ In any case, the spectrum at
scales below \lam\ will now contain glueballs, at least one of which will have
spin-0, and which  will be denoted by $G.$ As a result of (\ref{eq:sigf}),
there will
be $\sighg$ mixing, which can be parametrized by  a mixing term
\be
V_{mix}=-\mu\sigma G\ \ .
\label{eq:mix}
\ee
The coupling (\ref{eq:sigf}) and dimensional considerations imply
\be
\mu=\kappa\frac{\lam^3}{\mp}\ \ ,
\label{eq:mu}
\ee
with $\kappa\sim O(1).$ To complete the effective potential for the $\sighg$
system, I  introduce  glueball and dilaton mass terms, and some
quartic terms to stabilize the potential:
\be
V(\sigma,G)=\thalf\msigt\sigma^2+\thalf \mgt
G^2-\mu\ \sigg+\tquarter\lambda_{\sigma}\sigma^4+\tquarter\lambda_G G^4\ \ .
\label{eq:vsigg}
\ee
In Eq.~(\ref{eq:vsigg}),  $\msigt, \lsig$ and $\lg$ are unknown.
For the glueball mass, along with the estimates (\ref{eq:msig}) and
(\ref{eq:mu}), we expect
\be
m_G=\alpha \lam\ \ ,
\label{eq:mg}
\ee
with $\alpha\sim O(1).$ Although the parameters $\msigt,\lsig,\lg$ have been
introduced in complete
ignorance of their origin and magnitude, it will turn out, very remarkably,
that $(a)$ only $\lsig$ will be required to be non-zero and $(b)$
as long as $\lsig>O((\Lambda/\mp)^4),$ {\em the mass eigenvalues will be
determined entirely by $\mg$ and $\mu,$ and will not depend on
$\lsig$ to leading order in $\Lambda/\mp.$}\bigskip

\hedd{Mass Spectrum: Simplified Case}

I first note that, in the absence of the mixing term, the
potential has a  minimum at $\sigma=G=0.$ The mixing term shifts this
minimum, as well as the mass eigenvalues. We can obtain a general idea of the
result of this by considering a simplified situation, in which $\lg=\msigt=0.$
The potential then becomes
\be
\vzero\ (\sigma, G)=\thalf m_G^2
G^2-\mu\sigg+\tquarter\lambda_{\sigma}\sigma^4\
\ .
\label{eq:vzero}
\ee

The stationary conditions are
\begin{eqnarray}
\mu G- \lsig\sigma^3&=&0\nonumber\\
\mgt G-\mu\sigma&=&0\ \ .
\label{eq:stat}
\end{eqnarray}
These are consistent with either $\sigma=G=0,$ or with
\begin{eqnarray}
\bar\sigma&=&\pm\frac{1}{\sqrt{\lsig}}\left|\frac{\mu}{m_G}\right|\nonumber\\
\bar G&=&\pm\frac{1}{\sqrt{\lsig}}\frac{\mu}{\mgt}\left|\frac{\mu}{m_G}\right|
\ \ .
\label{eq:barsigg}
\end{eqnarray}
It is  simple to check that the zero-field point (0,0) is a saddle point
of $\vzero,$ whereas
$(\bar\sigma,\bar G)$ lies below that for the zero field values. Thus
$(\bar\sigma,\bar G)$ is a true minimum, with energy
\be
\vzero\ (\bar\sigma,\bar
G)=-\quarter\left(\frac{1}{\lsig}\right)\left(\frac{\mu}{m_G}\right)^4\ \ .
\label{eq:vbar}
\ee
It is important to note that as long as $\lsig\gg (\Lambda/\mp)^4,$ the shift
in $S$ due to the
new vacuum is $\delta S=\bar\sigma/\mp\ll 1/g^2.$

One may wish to constrain the parameters so that the
shift in  vacuum value of energy
%(before fine tuning in order to set the cosmological constant = 0)
be
$\lsim O(|F|^2,)$ where the $F$ is a SUSY-breaking
$F$-term, of $O(\mp\ms).$ From (\ref{eq:msig}), this is  equivalent to
requiring
\be
\left|\vzero\ (\bar\sigma,\bar G) - \vzero (0,0)\right|\lsim
\frac{\Lambda^6}{\mp^2}\ \ .
\label{eq:evac}
\ee
Eq.~(\ref{eq:evac}), combined with (\ref{eq:mu}), (\ref{eq:mg}), and
(\ref{eq:vbar}), implies \cite{lsig}
\be
\lsig\gsim O((\Lambda/\mp)^2)\ \ .
\label{eq:lsig}
\ee

Using (\ref{eq:barsigg}) and (\ref{eq:vzero}) one may  calculate the quadratic
fluctuation matrix ({\em i.e,} the $\sighg$
$(mass)^2$ matrix ${\cal M}^2)$ about
$(\bar\sigma,\bar G):$
\be
\left({\cal M}^2\right)_{\bar\sigma\bar G}=\left(
\begin{array}{cc}
3(\mu^2/\mgt)& -\mu\\[.02in]
-\mu&\mgt
\end{array}\right)
\label{eq:msq}
\ee

There is a remarkable aspect about this result: although a non-zero value of
$\lsig$ was required in order to establish the non-trivial vacuum, {\em the
fluctuations about this vacuum do not depend on the value of $\lsig.$} This
parameter has acted as a regulator, disappearing in the physical masses
(although not in the vacuum energy). The
eigenvalues may be immediately calculated, and are most transparent in leading
order in $(|\mu|/m_G^2)\sim \Lambda/\mp:$
\begin{eqnarray}
\msigp&\simeq&
\sqrt{2}\ \frac{|\mu|}{m_G}=
\sqrt{\frac{2\kappa}{\alpha}}\frac{\Lambda^2}{\mp}\nonumber\\[.1in]
m_{G\pri}&\simeq&m_G=\alpha\Lambda\ \ .
\label{eq:masses}
\end{eqnarray}\bigskip

\hedd{Mass Spectrum: General Case}

The treatment of the more general case is
slightly more complicated, but the result is the same.
In what follows, it will be convenient to set
$\mp=1$ until the end, and to define
\be
\epsilon\ \equiv \ \frac{\Lambda}{\mp}\ \ ,
\label{eq:eps}
\ee
In accordance with Eqs. (\ref{eq:mu}) and
(\ref{eq:mg}),
\be
\mu\sim \epsilon^3,\quad \mgt\sim \epsilon^2\ \ .
\label{eq:mepso}
\ee
Typically, models of dilaton potentials prior to mixing with glueball give
\be
\msigt\sim \ms^2\sim \epsilon^6\ \ .
\label{eq:mepst}
\ee
Beginning with the potential (\ref{eq:vsigg}), it is apparent that a tachyonic
instability develops in the quadratic sector if
\be
\msigt\ \mgt<\mu^2\ \  ,
\label{eq:tach}
\ee
which will certainly be the case with the estimates in Eqs.~(\ref{eq:mepso})
and (\ref{eq:mepst}). In such a case,  we search
for the displaced vacuum and spectrum for (\ref{eq:vsigg}). The extremum
conditions following from (\ref{eq:vsigg}) are
\begin{eqnarray}
\msigt\sigma-\mu G+\lsig\sigma^3 & = &0\nonumber\\
\mgt G-\mu \sigma + \lg G^3 &=&0\ \ .
\label{eq:statt}
\end{eqnarray}

These allow the solution $\sigma=G=0,$ with $V=0;\ $this is a saddle
point if (\ref{eq:tach}) is true. Alternately, there is a non-trivial solution
which follows from (\ref{eq:statt}). On eliminating $G$ via
\be
G=\frac{\sigma}{\mu}\ (\msigt+\lsig\sigma^2)\equiv
\frac{\sigma}{\mu}\cdot  \frac{\mu^2}{\mgt}\ x\ \ ,
\label{eq:gsx}
\ee
in terms of a dimensionless variable $x,$ one obtains from
Eq.~(\ref{eq:statt}) the quartic equation
\be
f(x)=\left(\frac{\mu}{\mgt}\right)^4\left(\frac{\lg}{\lsig}\right)
\left(x^4-\left(\frac{\msigt\mgt}{\mu^2}\right)x^3\right)+x-1=0\ \ .
\label{eq:x}
\ee

The analysis of the solution spectrum of (\ref{eq:x}) goes as follows:~
We first note that, in terms of the small quantity
$\epsilon$ (Eq.~(\ref{eq:eps})),
\begin{eqnarray}
\left(\frac{\mu}{\mgt}\right)^4&\sim& \epsilon^4\nonumber\\
\frac{\msigt\mgt}{\mu^2}&\sim& \epsilon^2\ \ .
\label{eq:othereps}
\end{eqnarray}
It is then easy to show that $(i)$  $f(0)<0,\ f(x\geq 1)>0;$~$(ii)$ if
$\lsig>O(\epsilon^{10}),$ then $f(x)$ is a monotonically increasing function
for $x\geq 0.$
In that case, there is one and only one root for positive $x,$ and
we can obtain  it by expanding around $x=1.$ After some algebra, one finds that
the stationary point occurs at
\be
\bar x\simeq 1+O(\epsilon^4)\ \ .
\label{eq:xbar}
\ee

After this, the analysis proceeds exactly as in the simplified case.
Once more, the stationary point $(\bar\sigma,\bar G)$ in field space
is given by Eq.~(\ref{eq:barsigg}) up to small corrections, and the energy
density at the true vacuum
$(\bar\sigma,\bar G)$ is given (up to corrections of $O(\epsilon^4))$ by
(\ref{eq:vbar}). Thus, in the general case, the energy density at
$(\bar\sigma,\bar G)$ is lower than at  $\sigma=G=0.$
The masses (to leading order in
$\epsilon)$ are as before:
\setcounter{equation}{14}
\begin{eqnarray}
\msigp&\simeq&
\sqrt{2}\ \frac{|\mu|}{m_G}=
\sqrt{\frac{2\kappa}{\alpha}}\frac{\Lambda^2}{\mp}\nonumber\\[.1in]
m_{G\pri}&\simeq&m_G=\alpha\Lambda\ \ .
\end{eqnarray}
The result of all this is a dilaton mass
\setcounter{equation}{24}
\be
m_{dilaton}=\msigp\sim (\ms^2\mp)^{\frac{1}{3}}\simeq 10^8
\mbox{GeV}\ \ .
\label{eq:mdil}
\ee
\bigskip

\hedd{The Decay of the Dilaton}

{}From the mass matrix (\ref{eq:msq}) and the eigenvalues (\ref{eq:masses}), it
is seen that the dilaton eigenstate $\sigma\pri$ differs from the uncoupled
$\sigma$ by terms of $O(\epsilon):$
\be
\sigma\pri\simeq \sigma + (\mu/\mgt)G\ \ ,
\label{eq:sigpsig}
\ee
so that the coupling of the dilaton to standard model gauge fields can be
obtained from (\ref{eq:sigf}). This immediately yields a lifetime
\begin{eqnarray}
\tau_{dilaton}\approx\frac{\mp^2}{\msigp^3}&\approx
&\frac{\mp^5}{\Lambda^6}\nonumber\\ &\approx& \frac{\mp}{\ms^2}\ \ ,
\label{eq:td}
\end{eqnarray}
so that the decay of the dilaton precedes (or is coincident with) the
electroweak era \cite{glueball} .Thus, even if a large amount of energy
resides in the
oscillations of the dilaton field about its minimum, the reheating caused
by its decay would not endanger nucleosynthesis
\cite{dilatondecay} or even anomalous baryosynthesis
\cite{kuzmin},\cite{hg}.\bigskip

\hedd{Conclusions}

It has been shown that the mixing of a dilaton with a heavy scalar
glueball, asdetermined by the canonical coupling of the dilaton, will
result in a dilatonof mass $\sim \Lambda^2/\mp,$ where $\Lambda$  is the
confinement scale  of the gauge theory. The result is due to a specific
interplay of the mass of the glueball and  the size of the coupling.
A quartic stabilization of the dilaton potential is necessary for the
dynamics, but the magnitude of the coupling
plays no role (to leading order in $\Lambda/\mp)$ in
determining the mass or the
coupling of the physical dilaton. Finally, the large mass of the dilaton is
shown to circumvent the usual cosmological problems associated with its
existence.
\section*{Acknowledgment}
I would like to thank Tom Taylor for many helpful conversations
concerning recent work on string-based SUSY-breaking mechanism. A discussion
with Mike Vaughn concerning the analysis of Eq.~(\ref{eq:x}) is appreciated.
This research was supported in part
by the National Science Foundation (Grant No. PHY-9306524), and by the
Texas National Research Laboratory Commission (Award No. RGFY93-277).\clearpage

\begin{thebibliography}{99}
\bibitem{witten}Aspects of the dilaton are reviewed in M.~B.~Green,
J.~H.~Schwartz, and E.~Witten, {\em Superstring Theory,} Cambridge University
Press, 1987.
\bibitem{dilpot}M.~Dine and N.~Seiberg, \plett{162}299 (1985);  for
recent efforts in the context of the gaugino-dilaton-moduli framework
of supersymmetry-breaking, see \nl S.~Ferrara, N.~Magnoli, T.~R.~Taylor, and
G.~Veneziano,
\plett{245}243 (1990);\nl
 A.~Font, L.~Ib\`a\~nez, D.~Lust, and F.~Quevedo,
\plett{245}401 (1990);\nl
H.~P.~Nilles and M.~Olechowski, \plett{248}268 (1990);\nl
P.~Binetruy and M.~K.~Gaillard, \plett{253}119 (1991); \nl A.~de~la~Macorra and
G.~G.~Ross, \np{404}321 (1993); \nl B.~de Carlos, J.~A.~Casas, and C.~Mu\~noz,
\np{399}623 (1993). Recent discussions of
the cosmological problems associated with the dilaton potential are given by
\nl R.~Brustein and P.~J.~Steinhardt, \plett{302}196 (1993);
\nl B.~de Carlos, J.~A.~Casas, F.~Quevedo, and E.~Roulet, \plett{318}447
(1993).
\bibitem{dilatondecay}G.~D.~Coughlan, W.~Fischler, E.~W.Kolb, S.~Raby, and
G.~G.~Ross, \plett{131}59, (1983);\nl
G.~German and G.~G.~Ross, {\em ibid\ } {\bf 172}, 305 (1986);\nl  J.~Ellis,
D.~V.~Nanopoulos, and M.~Quiros, {\em ibid\ } {\bf 174}, 176 (1986);\nl
O.~Bertolami, {\em ibid\ } {\bf 209}, 277 (1988).  See also T.~Banks,
D.~B.~Kaplan, and A.~E.~Nelson, hep-ph/9308292, and B.~de~Carlos {\em et~al,}
\cite{dilpot}, for   recent discussion.
\bibitem{nillesetal}S.~Ferrara, L.~Girardello, and H.~P.~Nilles, \plett{125}457
(1983); \nl J.~-~P.~Derendinger, L.~E.~Ib\`a\~nez, and H.~P.~Nilles,
{\em ibid\ }
{\bf 155}, 65 (1985);\nl M.~Dine, R.~Rohm, N.~Seiberg, and E.~Witten,
{\em  ibid\ } {\bf 156}, 55 (1985);\nl C.~Kounnas and M.~Porrati,
{\em ibid\ } {\bf 191}, 91
(1987). For more recent discussion, see works in Ref.~\cite{dilpot}.
\bibitem{lsig}The value of $\lsig$ is highly model dependent. In general,
multiple gaugino condensates are required in order to yield a positive $\lsig$
near the local minimum of the dilaton potential. This can be seen, {\it e.g.,}
by examining some of the models presented in  the paper of B.~de~Carlos {\it et
al,} Ref.\cite{dilpot}.
\bibitem{glueball}The lifetime of the glueball is shorter by a factor
$\Lambda/\mp.$
\bibitem{kuzmin}V.~Kuzmin, V.~Rubakov, and M.~Shaposhnikov, \plett{155}36
(1985); P.~Arnold and McLerran, \prd{36}581 (1987).
\bibitem{hg}A detailed discussion of the condensate field
as inflaton, and of the reheating effects of dilaton decay,
have been given by the author in another publication
(H.~Goldberg, Northeastern University preprint NUB-3082/93-Th, hep-ph/9312282)
\eb\ed